\documentclass[aps,showpacs,twocolumn,superscriptaddress]{revtex4}
\usepackage{amsfonts}
\usepackage{amsmath}
\usepackage{amssymb}
\usepackage{graphicx}
\usepackage{bm}

\begin{document}

\title{Flow pattern in the vicinity of self-propelling hot Janus particles}
\author{Thomas Bickel}
\affiliation{LOMA, Universit\'{e} de Bordeaux \& CNRS, 351 cours de la Lib\'{e}ration,
33405 Talence, France}
\author{Arghya Majee}
\affiliation{LOMA, Universit\'{e} de Bordeaux \& CNRS, 351 cours de la Lib\'{e}ration,
33405 Talence, France}
\affiliation{Max-Planck-Institut f\"ur Intelligente Systeme, Heisenbergstra\ss e 3, 70569
Stuttgart, Germany}
\author{Alois W\"urger}
\affiliation{LOMA, Universit\'{e} de Bordeaux \& CNRS, 351 cours de la Lib\'{e}ration,
33405 Talence, France}
\affiliation{ZHS, Universit\"at Leipzig, Burgstra\ss e 21, 04103 Leipzig, Germany}

\begin{abstract}
We study the temperature field and the resulting flow pattern in the
vicinity of a heated metal-capped Janus particle. If its thickness exceeds
about ten nanometers, the cap forms an isotherm and the flow pattern
comprises a quadrupolar term that decays with the square of the inverse
distance~$\sim r^{-2}$. For much thinner caps the velocity varies as~$\sim
r^{-3}$. These findings could be relevant for collective effects in dense
suspensions and for the circular tracer motion observed recently in the
vicinity of a tethered Janus particle.
\end{abstract}

\pacs{82.70.Dd, 66.10.cd,47.15.G-}
\maketitle

\textit{Introduction.} The design of artificial micro- and nano-swimmers
that propel themselves in a viscous fluid is a key issue in nanotechnology~%
\cite{Ebb10}. In the realm of biology, autonomous motion of microorganisms
is ubiquitous and relies on surface waves or periodic body deformations~\cite%
{Lau09}. Several swimming devices inspired by living systems have been built
recently~\cite{Dre05,Gos09}, although their actuation mechanism requires
external forces or torques. An alternative way toward self-propulsion is
achieved using colloidal particles with non-uniform surface properties~\cite%
{Gol05}. This class of rigid swimmers relies on phoretic transport, \textit{%
i.e.} the force-free motion driven by the gradient of an external field~\cite%
{And89}. In the case of self-phoresis, however, asymmetric particles are
able to generate their own gradient within an otherwise homogeneous medium
and thus to convert the available energy into mechanical work~\cite%
{Gol07,Jue09}.

The simplest realization of autonomous swimmers is obtained with Janus
particles, which are colloidal objects with two sides differing in their
physical or chemical properties~\cite{Wal08}. For example, a bimetallic
particle in a peroxide solution generates a electrochemical gradient which
in turn gives rise to a flow in the surrounding fluid and thus causes
self-propulsion~\cite{Pax05}. At short times this results in linear motion,
whereas at longer times the random reorientations lead to enhanced diffusion~%
\cite{How07,Dun12}. Similar findings have been reported for photophoresis of
hot Janus particles, which move in their own temperature gradient, with an
effective diffusion coefficient that increases linearly with the heating
power~\cite{Jia10,Vol11,But12}.

Heating of metal capped Janus particles provides a versatile means of
actuation which, in particular, can be switched on and off almost
instantaneously. Heat absorption of the metal cap is achieved upon
illumination by a defocused laser beam~\cite{Jia10,Vol11,But12,Qia13} or
when subject to an ac magnetic field~\cite{Bar13}. The metal patch absorbs
the energy and converts it into heat; asymmetric heat release then drives
the colloid via a mechanism of thermophoresis~\cite{And89}.

\begin{figure}[b]
\includegraphics[width=\columnwidth]{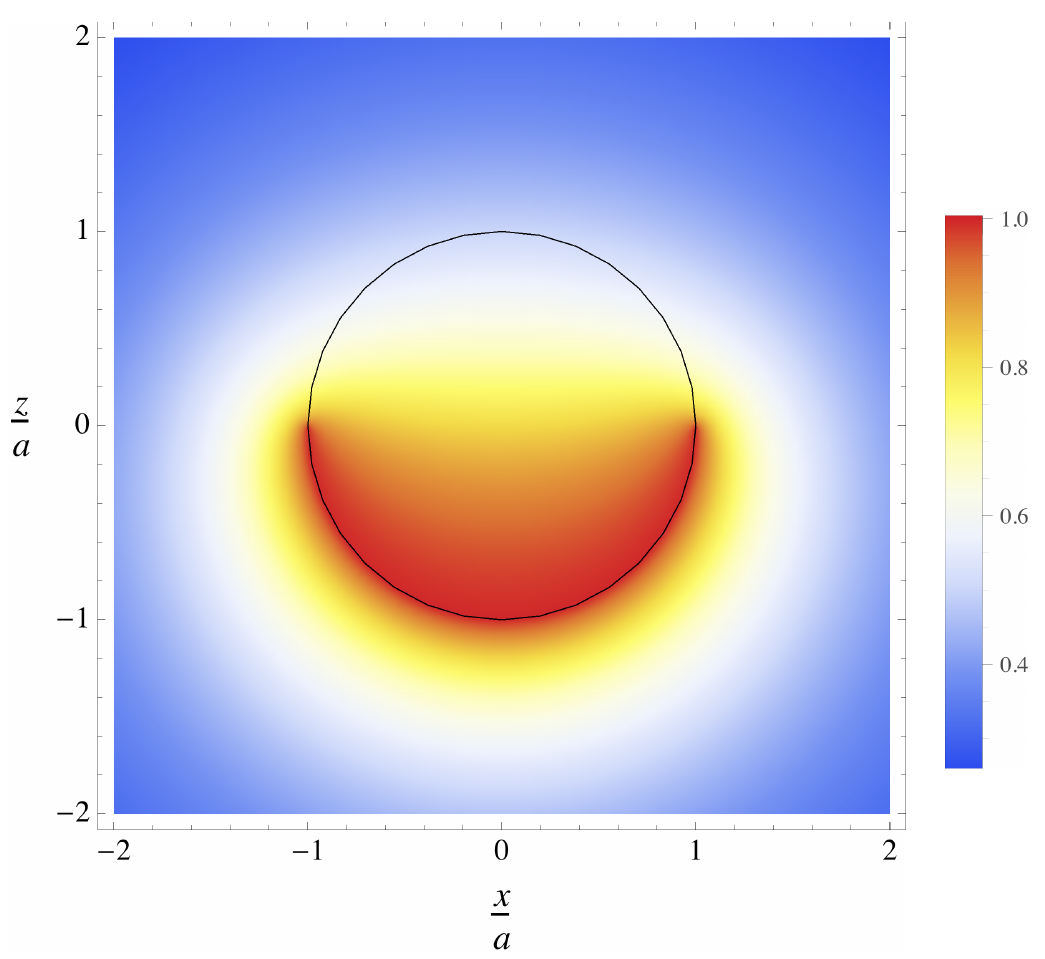}
\caption{(Color online) Map of the reduced temperature field $%
(T-T_{0})/\Delta T$ in the $(xOz)$ plane, inside and outside the
metal-coated colloid.}
\label{fig1}
\end{figure}

In this Letter we address the temperature profile and the fluid velocity in
the vicinity of a hot Janus particle. In view of its large thermal
conductivity we treat the metal cap as an isotherm and, as a consequence,
obtain a class of hydrodynamic multipoles that are absent when neglecting
heat conduction in the cap. These additionall terms result in a flow pattern
which is strongly asymmetric with respect to the particle midplane, and
could affect the hydrodynamic coupling between neighboring swimmers or with
a bounding wall~\cite{Ish06,Llo10,Gol12}. On the other hand, a fixed Janus
particle is expected to act as a micro-pump. Visualization of the local
convective flow by particle tracking velocimetry indeed revealed vortices
close to a Janus particle tethered on a glass support~\cite{Jia10}. The
description of tracers trajectories thus requires a detailed knowledge of
both the temperature and the velocity fields around a Janus particle.

\textit{Temperature field.} In a first step we derive the temperature
profile from Fourier's law 
\begin{equation}
\kappa \mathbf{\nabla }^{2}T=q(\mathbf{r})\ ,  \label{fourier}
\end{equation}%
where $q$ is the power absorbed by the metal cap, and $\kappa $ the thermal
conductivity of both the particle and the surrounding fluid (taken to be the
same for simplicity.) If cap conductivity $\kappa _{c}$ is usually much
higher than $\kappa $; if their ratio is larger than the ratio of particle
radius and cap thickness,  $\kappa _{c}/\kappa >a/d$, the cap forms an
isotherm. Since this condition is satisfied for a 50 nm gold cap on
micron-size silica or polystyrene beads, we assume in the following a
constant cap temperature $T_{0}+\Delta T$. Typical values for the excess
temperature $\Delta T$ with respect to the bulk are of the order of a few
Kelvins. 

Because of the mixed boundary conditions, constant temperature on the metal
cap and heat flux continuity on the upper hemisphere, there is no
straightforward solution of Eq.~(\ref{fourier}). As shown in the
supplementary material  \cite{supplmat}, the global constraint on the
isotherm can be implemented by a method based on auxiliary functions.  Here
we merely quote the temperature profile in the liquid phase ($r>a$) 
\begin{subequations}
\label{tempfield}
\begin{equation}
T(r,\theta )=T_{0}+\frac{\Delta T}{\pi }\sum_{n=0}^{\infty
}t_{n}P_{n}(c)\left( \frac{a}{r}\right) ^{n+1}\ ,
\end{equation}%
with $c=\cos \theta $ and the Legendre polynomial $P_{n}$. The coefficients $%
t_{n}$ are given by 
\end{subequations}
\begin{equation}
t_{2k}=-t_{2k+1}=\frac{(-1)^{k}}{2k+1}\ ,  \label{3}
\end{equation}%
except for the first one that reads $t_{0}=1+\pi /2$.\ A similar expression
with the same coefficients is found inside the particle ($r<a$), albeit with 
$(r/a)^{n}$ instead of $\left( a/r\right) ^{n+1}$. Identifying the power $%
\mathcal{P}$ absorbed by the metal cap with the total outward heat flow, one
readily establishes the relation with the excess temperature: $\mathcal{P}%
=(2\pi +4)\kappa a\Delta T$, with $\kappa $ the thermal conductivity of the
liquid. The map of the temperature field is shown in Fig.~\ref{fig1}. The
role of the isotherm assumption is illustrated by comparing with the case of
a very thin cap where $\kappa _{c}/\kappa <a/d$. Then the heat conductivity
of the metal structure can be neglected, and one readily finds that the even
coefficients of the temperature profile vanish, $t_{2k}=0$ \cite{Jia10}.   

\begin{figure}[tbp]
\includegraphics[width=\columnwidth]{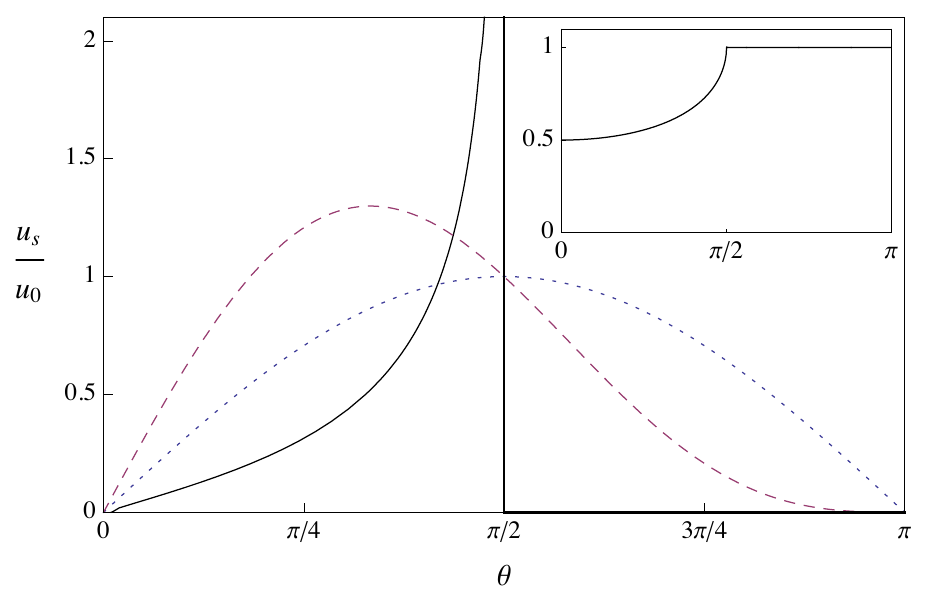}
\caption{(Color online) Quasi-slip velocity $u_s(\protect\theta)$ as a
function of the polar angle $\protect\theta$. For comparison, we plot the
infinite series~(\protect\ref{slipvelocity}) (solid line), the dipolar
approximation (series truncated at $n=1$, dotted line) and the quadrupolar
approximation (series truncated at $n=2$, dashed line). Inset: surface
temperature $[T(a,\protect\theta)-T_{0}]/\Delta T$ as a function of $\protect%
\theta$.}
\label{fig2}
\end{figure}

\textit{Boundary velocity.} The temperature gradient modifies the
particle-solvent interactions in a boundary layer of thickness $\ell $. For
electric-double layer forces $\ell $ is given by the Debye length, and for
depletion forces by the gyration radius of the polymers. In both cases $\ell 
$ is much smaller than the radius $a$ of micron size colloidal particles,
such that the flow pattern in the liquid can be evaluated in boundary layer
approximation~\cite{And89,Fay08,Wue10}. The excess enthalpy density $h$
results in a quasi-slip velocity of the liquid with respect to the particle~%
\cite{Der87}.

The boundary velocity is proportional to the temperature gradient parallel
to the surface of the particle $u_{s}=-(\ell ^{2}\bar{h}/\eta T_{0})\nabla
T_{||}$, where $\eta $ is the viscosity and $\bar{h}$ the characteristic
value of the excess enthalpy. With Eq.~(\ref{tempfield}) one has 
\begin{equation}
u_{s}(\theta )=u_{0}\sum_{n=1}^{\infty }t_{n}\frac{dP_{n}(c)}{d\theta } \ ,
\label{slipvelocity}
\end{equation}
where the prefactor $u_{0}$ gives the velocity scale, 
\begin{equation}
u_{0}=-\frac{\ell ^{2}\bar{h}}{\pi \eta a}\frac{\Delta T}{T_{0}} \ .
\label{defu0}
\end{equation}
The first term in Eq.~(\ref{slipvelocity}) corresponds to the dipolar
approximation: $u_{s}(\theta )=u_{0}\sin \theta $~\cite{And89}. Keeping the
first two terms of the series, the surface velocity is that of a
``squirmer'' with positive stresslet $\beta=2/3$~\cite{Bla71,Llo10}.

For positive slip velocity $u_{0}$, i.e., negative enthalpy $\bar{h}$, the
liquid flows toward the warmer side of the Janus particle. Note that $u_{s}$
is largest on the upper half-sphere close to mid-plane; it vanishes on the
lower half-sphere because of the constant temperature of the metal cap --
see Fig.~\ref{fig2}. The expression $\ell ^{2}\bar{h}$ has the dimension of
a force, and has been evaluated for several thermophoretic mechanisms.
Ruckenstein pointed out the positive slip velocity ($u_{0}>0$) due to the
enthalpy of the electric double layer, $\ell ^{2}\bar{h}=-\frac{1}{2}%
\varepsilon \zeta ^{2}$~\cite{Ruc81}, with the surface potential $\zeta $
and the solvent permittivity $\varepsilon $.\ In many instances, however,
the slip velocity is dominated by the thermoelectric effect $\ell ^{2}\bar{h}%
=\frac{3}{2}\varepsilon \zeta ST_{0}$, where the electrolyte Seebeck
coefficient $S$ may take either sign~\cite{Put05,Wue08,Vig10}. Upon adding
polymer to the solution, thermal depletion forces result in $u_{0}<0$~\cite%
{Jia09}. For a micro-size particle with $\Delta T=1$ K, the slip velocity is
a few microns per second.

\textit{Bulk velocity field. }The quasi-slip velocity on the surface of the
particle induces a flow in the surrounding liquid. The general axisymmetric
solution $\mathbf{v=}v_{r}\mathbf{e}_{r}+v_{\theta }\mathbf{e}_{\theta }$ of
the Stokes' equation has been known for a long time~\cite{Bre61,Bla71}. Here
we give the series expansion of Ref.~\cite{Morth10}, where the radial and
tangential components are given by 
\begin{subequations}
\label{velocity}
\begin{align}
v_{r}=u_{0}\sum_{n=1}^{\infty }\frac{a^{n}}{r^{n}}\left( p_{n}+q_{n+2}\frac{%
a^{2}}{r^{2}}\right) P_{n}(c) \ ,  \label{vr} \\
v_{\theta }=u_{0}s\sum_{n=1}^{\infty }\frac{a^{n}}{r^{n}}\left( p_{n}\frac{%
n-2}{n(n+1)}+\frac{q_{n+2}}{n+1}\frac{a^{2}}{r^{2}}\right) P_{n}^{\prime
}(c) \ ,  \label{vt}
\end{align}
with $P_{n}^{\prime }=dP_{n}/dc$ and $s=\sin \theta $. The corresponding
pressure field is given in~\cite{supplmat}. The coefficients $p_{n}$
describe the inhomogeneous solutions of Stokes' equation with finite
pressure, whereas the $q_{n}$'s are related to the zero-pressure homogeneous
solutions. The coefficients are set by the boundary conditions at the
surface of the particle. First, the far field $\mathbf{v}$ has to match the
sum of the particle velocity $u_{p}\mathbf{e}_{z}$ and the quasi-slip
velocity 
\end{subequations}
\begin{equation}
\mathbf{v}|_{r=a}=u_{p}\mathbf{e}_{z}+u_{s}\mathbf{e}_{\theta } \ .
\label{14}
\end{equation}
The second condition is a global constraint and involves the total force $%
F_{z}=-4\pi \eta u_{0}ap_{1}$. In the following we evaluate the coefficients
for a particle that is either freely moving of fixed at a given position.

\begin{figure}[tbp]
\includegraphics[width=\columnwidth]{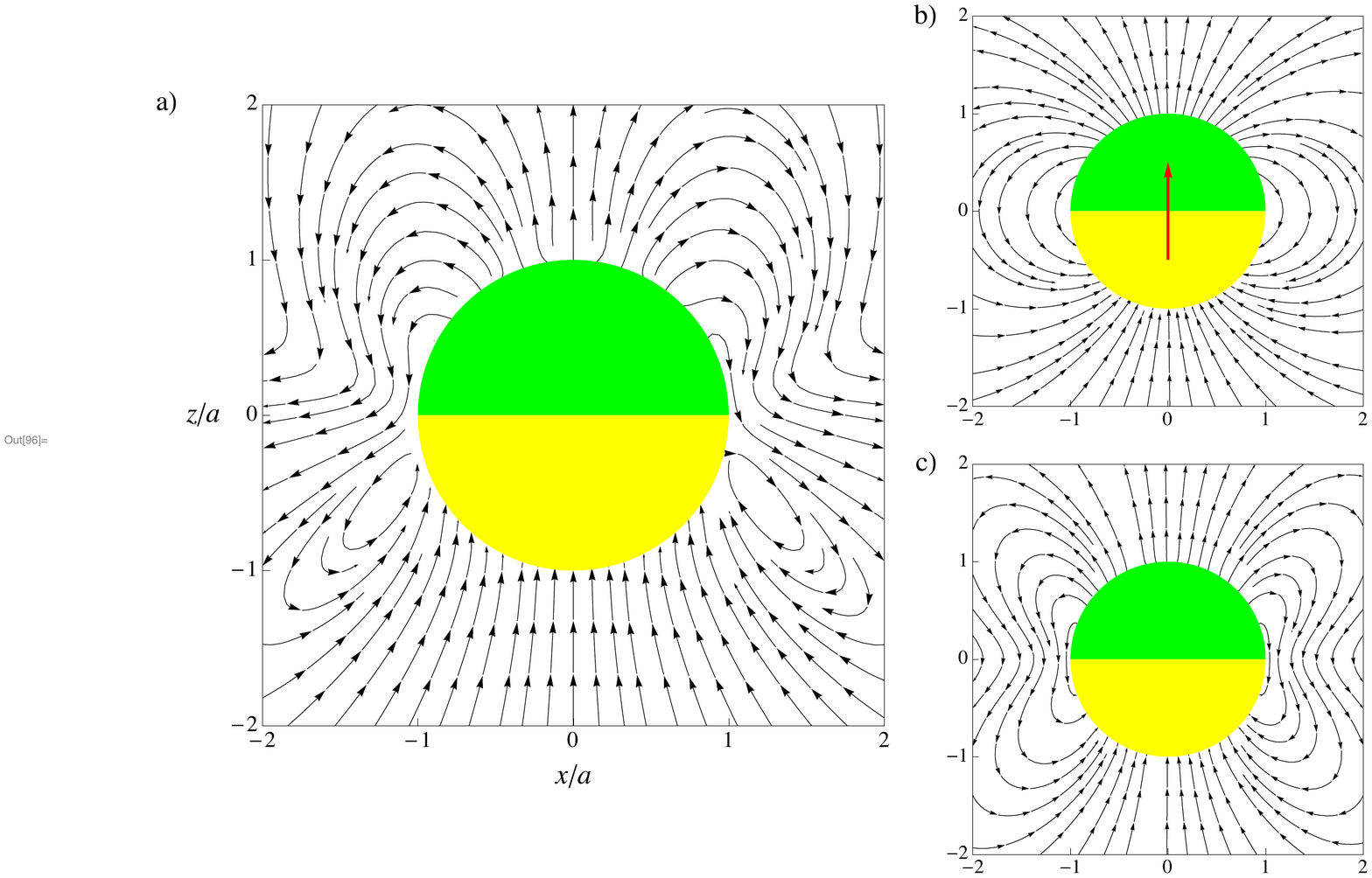}
\caption{(Color online) Flow streamlines around a moving Janus particle in
the laboratory frame. a) The cap forms an isotherm with $t_{k}$ as in (%
\protect\ref{3}). b) Dipolar approximation with $t_{1}$ only. c) Thin-cap
limit, with $t_{2k}=0.$}
\label{fig3}
\end{figure}

\textit{Moving particle.} First we consider a free Janus particle that
self-propels due its own temperature gradient. Since there is no external
force, the global constraint imposes the well-known condition $p_{1}=0$~\cite%
{And89}. Yet, the no-force condition does not affect the inhomogeneous
coefficients of higher order. Noting $\mathbf{e}_{z}=c\,\mathbf{e}_{r}-s\,%
\mathbf{e}_{\theta }$, one obtains the radial and tangential projections of
Eq.~(\ref{14}), $v_{r}=cu_{p}\ $and $v_{\theta }=-su_{p}+u_{s}.$ Inserting
Eqs.~(\ref{vr}) and~(\ref{vt}), one readily gets for $n=1$ 
\begin{equation}
p_{1}=0\ ,\ \text{and}\quad q_{3}=-\frac{2}{3}t_{1}=\frac{2}{3} \ ,
\label{coefmob1}
\end{equation}
and for higher orders 
\begin{equation}
p_{n}=-q_{n+2}=\frac{n(n+1)}{2}t_{n}\qquad (n\geq 2) \ .  \label{coefmob}
\end{equation}
The particle velocity is then in opposite direction to the quasi-slip and is
equal to two thirds of its amplitude 
\begin{equation}
u_{p}=q_{3}u_{0}=\frac{2}{3}u_{0} \ .  \label{32}
\end{equation}
This implies that self-propulsion is driven by the dipolar term $q_{3}$
only; higher Fourier coefficients of the temperature gradient, $t_{n}$ with $%
n>1$, do not contribute to the particle velocity.

\begin{figure}[tbp]
\includegraphics[width=\columnwidth]{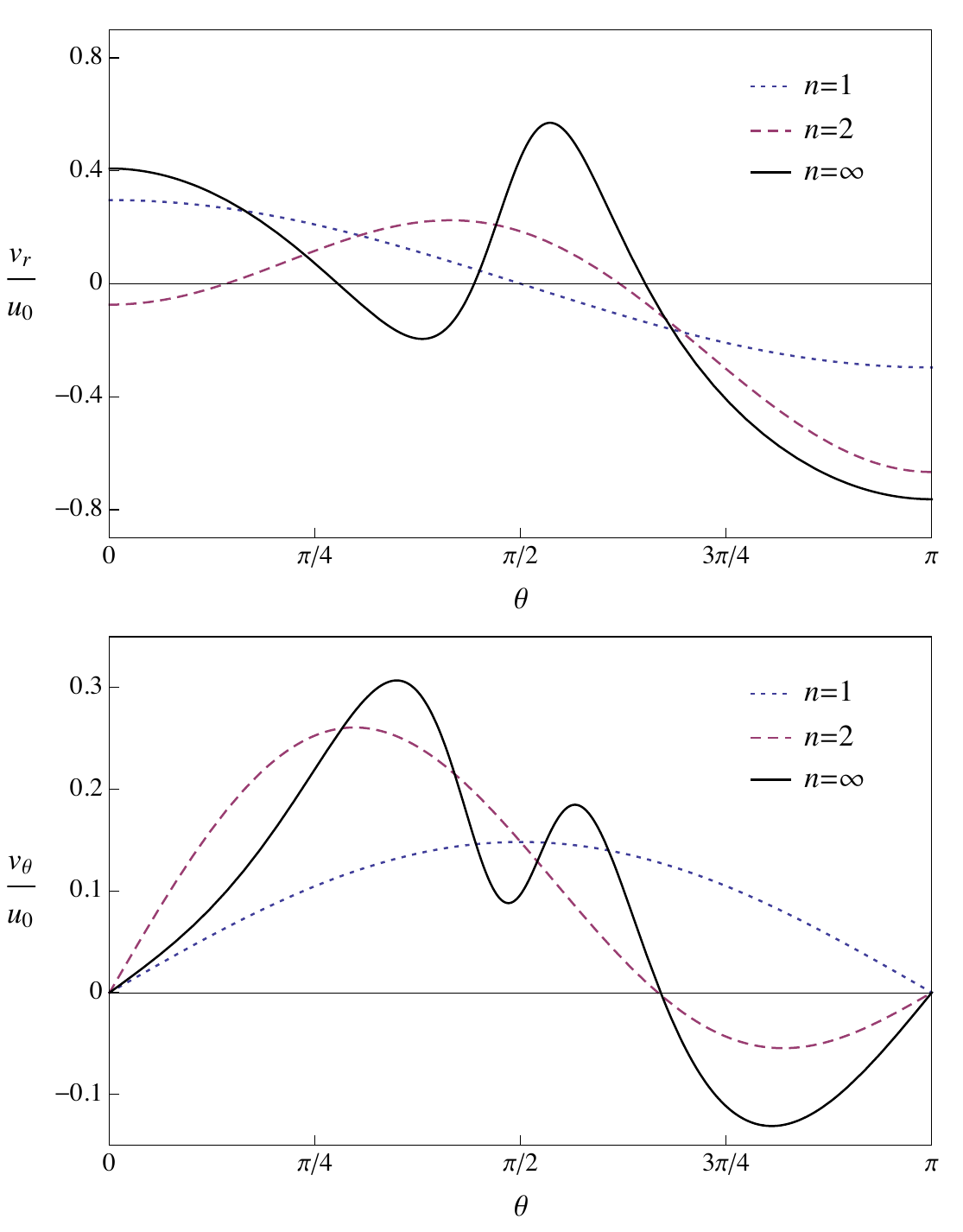}
\caption{(Color online) Radial and tangential components of the velocity of
a moving particle at distance $r=1.5a$ from its center. The plots correspond
to the series~(\protect\ref{vr}) and~(\protect\ref{vt}) truncated at $n=1$
(dipolar approximation, dotted lines), truncated at $n=2$ (quadrupolar
approximation, dashed lines), and to the infinite series (solid lines).}
\label{fig4}
\end{figure}

We emphasize two major differences with respect to the thin-cap limit, where 
$t_{2n}=0$ and the dipolar approximation where $t_{1}$ is the only non-zero
coefficient. First, the temperature coefficient  $t_{2}$ results in a radial
velocity contribution that decays with the square of the inverse distance
and shows quadrupole characteristics. In the thin-cap and dipolar
approximations the velocity decays as $r^{-3}$. Second, Fig.~\ref{fig3}a)
shows that the rotational patterns of the stream lines are located close to
the metal cap; for comparison, we also plot the dipolar flow field with the
only coefficient $q_{3}$ and the thin-cap limit with $t_{2k}=0$. The
corresponding streamlines in Fig.~\ref{fig3}b) and~c) are symmetric with
respect to midplane.

In Fig.~\ref{fig4} we plot both $v_{r}$ and $v_{\theta }$ as a function of $%
\theta $ at a distance $r=1.5a$ from the center of the particle; we compare
the whole series with the dipolar approximation ($n=1$ only), and the
quadrupolar approximation ($n=1,2$). The dipolar terms are simply given by
sine and cosine functions. The quadrupolar correction is by no means small
or insignificant; for example, $v_{r}$ changes sign at small $\theta $, and $%
v_{\theta }$ at angles close to $\pi $. Retaining the higher-order
corrections again changes the flow pattern drastically. As the most striking
feature, note the large positive derivative $dv_{r}/d\theta $ close to
midplane; together with the positive value of the tangential component $%
v_{\theta }$ this implies the existence of vortices at the edge of the metal
cap.

\begin{figure}[t]
\includegraphics[width=\columnwidth]{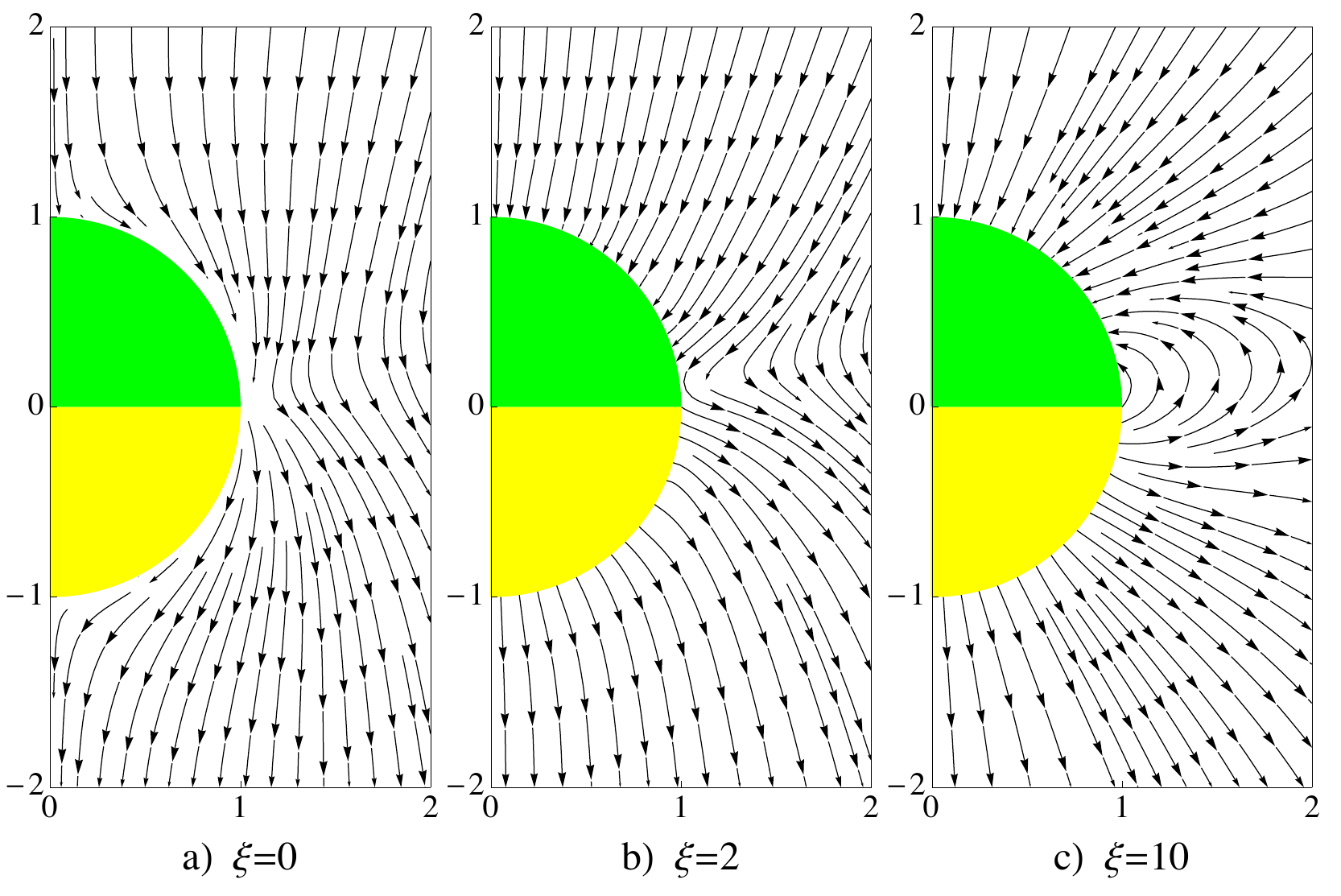}
\caption{(Color online) Map of the tracer velocity $\mathbf{u}_{t}$ for
different values of the parameter~$\protect\xi $.}
\label{fig5}
\end{figure}

\textit{Immobile particle. } Now we turn to the situation where the Janus
particle is fixed at a given position. This requires a finite external force
that counteracts the self-propelling surface stress in the boundary layer.
The particle velocity is then zero, $u_{p}=0$, so that the quasislip
velocity matches to the tangential component of the far-field $u_{\theta
}=v_{\theta }$, whereas the radial component vanishes $v_{r}=0$. One obtains
the coefficients for the flow pattern 
\begin{equation}
p_{n}=-q_{n+2}=\frac{n(n+1)}{2}t_{n} \qquad (n\geq 1) \ .  \label{coefim}
\end{equation}
With the coefficient $p_{1}=-1$ one finds the external force $F_{z}=4\pi
\eta au_{0}$.\ It is required to immobilize the particle which otherwise
would move at a velocity $u_{p}=\frac{2}{3}u_{0}$, and thus corresponds to
the well-known Stokes drag $6\pi \eta au_{p}$.

In Fig.~\ref{fig5}a) we plot the flow pattern $\mathbf{v}(r,\theta )$.
Contrary to that of the moving particle, there are no vortices close to the
particle; the liquid flows smoothly around the immobile particle. Comparison
of the coefficients shows that this difference is only due to the
lowest-order coefficients $p_{1}=-1$ and $q_{3}=1$; in other words, the
large long-range contribution $p_{1}$ hides the vortices that accompany a
moving particle but are invisible in the case where the particle is fixed.

\textit{Motion of a tracer particle.} Finally, we consider a small tracer in
the neighborhood of a fixed Janus particle. Its velocity $\mathbf{u}_{t}$ is
given by the sum of the convective flow and of thermophoretic drift in the
temperature gradient of the Janus particle 
\begin{equation}
\mathbf{u}_{t}=\mathbf{v}(\mathbf{r})-D_{T}\nabla T \ ,  \label{transport}
\end{equation}
with $\mathbf{v}(\mathbf{r})$ given by Eqs.~(\ref{velocity}) and~(\ref%
{coefim}). The mobility coefficient $D_T$ is expressed by the enthalpy
density and thickness of the boundary layer of the tracer~\cite{Wue10}. The
vector fields $\mathbf{v}$\ and $\bm{\nabla} T$ have different
characteristics: at large distance, the first is isotropic and the second
one of quadrupolar symmetry; close to the particle higher order terms lead
to an even more intricate variation. The relative importance of the two
terms in~(\ref{transport}) is expressed by the mobility ratio of tracer and
Janus particle, $\xi = D_T/{\hat D}_T$, which depends on their surface
properties. Either term in Eq.~(\ref{transport}) may be dominant, and they
may even carry opposite sign.

In Fig.~\ref{fig5} we plot the tracer velocity $\mathbf{u}_{t}$ for $\xi =0$%
, 2 and 10. As the most striking feature, the tracer is pushed toward the
colder half of the Janus particle from above but are strongly repelled from
the warmer side. For $\xi =0$ (no thermophoresis), the tracer first flows
toward the Janus particle, then creeps slowly toward the metal cap, and
finally is repelled from it. For intermediate value $\xi=2$, transport
alongside the surface has ceased and tracer particles either accumulate at
the upper side or are pushed away from the lower side. For the larger value $%
\xi =10$, the flow pattern shows additional vortices close to the midplane
of the Janus particle, so that tracers are brought back to the colder side.

Experimentally, flow circulation around a heated Janus particle tethered to
a glass surface has been reported recently~\cite{Jia10}. Tracking of
fluorescent particles moreover revealed that the concentration of tracers is
higher on the non coated side and lower on the coated side. It is thus
likely that the observed flow pattern results from the competition between
convection and thermophoresis, as expressed by Eq.~(\ref{transport}).

In summary, we have characterized the flow around a heated Janus colloid.
The discontinuity of surface properties has a major impact on the fluid
velocity field not only in the vicinity of the particle but also in the
bulk. In particular, we have shown that the dipolar approximation which is
usually considered for simplicity is only a poor approximation of the full
series. Taking into account higher order terms leads to a complex flow field
that can be relevant at finite concentration, where collective effects come
into play~\cite{Gol12,The12}.


\begin{thebibliography}{99}
\bibitem{Ebb10} S.J. Ebbens and J.R. Howse, Soft Matter \textbf{6}, 726--738
(2010)

\bibitem{Lau09} E. Lauga and T.R. Powers, Rep. Prog. Phys. \textbf{72},
096601 (2009)

\bibitem{Dre05} R. Dreyfus, J. Baudry, M.L. Roper, M. Fermigier, H.A. Stone,
and J. Bibette, Nature \textbf{437}, 862 (2005)

\bibitem{Gos09} A. Gosh and P. Fisher, Nano Lett. \textbf{9}, 2243 (2009)

\bibitem{Gol05} R. Golestanian, T.B. Liverpool, and A. Ajdari, Phys. Rev.
Lett. \textbf{94}, 220801 (2005)

\bibitem{And89} J.O. Anderson, Ann. Rev. Fluid Mech. \textbf{21}, 61 (1989)

\bibitem{Gol07} R. Golestanian, T.B. Liverpool, and A. Ajdari, New J. Phys. 
\textbf{9}, 126 (2007)

\bibitem{Jue09} F.\ Julicher, J. Prost, Eur. Phys.\ J. E \textbf{29}, 27
(2009)

\bibitem{Wal08} A. Walther and A.H.E. M\"uller, Soft Matter \textbf{4},
663--668 (2008)

\bibitem{Pax05} W. F. Paxton, A. Sen, and T. E. Mallouk, Chem. Eur. J. 
\textbf{11}, 6462 (2005).

\bibitem{How07} J.R. Howse, R.A.L. Jones, A.J. Ryan, T. Gough, R.
Vafabakhsh, and R. Golestanian, Phys. Rev. Lett. \textbf{99}, 048102 (2007)

\bibitem{Dun12} G. Dunderdale, S. Ebbens, P. Fairclough, and J. Howse,
Langmuir \textbf{28}, 10997 (2012)

\bibitem{Jia10} H.-R. Jiang, N. Yoshinaga, M. Sano, Phys. Rev. Lett. \textbf{%
105}, 268302 (2010)

\bibitem{Vol11} G. Volpe, I. Buttinoni, D. Vogt, H.-J. K\"ummerer, and C.
Bechinger, Soft Matter \textbf{7}, 8810 (2011)

\bibitem{But12} I. Buttinoni, G. Volpe, F. K\"{u}mmel, G. Volpe, C.
Bechinger, J. Phys.: Cond. Mat. \textbf{24}, 284129 (2012)

\bibitem{Qia13} B. Qian, D. Montiel, A. Bregulla, F. Cichos, H. Yang, Chem.
Sci. \textbf{4}, 1420 (2013)

\bibitem{Bar13} L. Baraban, R. Streubel, D. Makarov, L., D. Karnaushenko,
O.G. Schmidt, G. Cuniberti, ACS Nano \textbf{7}, 1360 (2013)

\bibitem{Ish06} T. Ishikawa, M. Simmonds, T. Pedley, J. Fluid Mech. \textbf{%
568}, 119 (2006)

\bibitem{Llo10} I. Llopis and I. Pagonabarraga, J. Non-Newtonian Fluid Mech. 
\textbf{165}, 946 (2010).

\bibitem{Gol12} R. Golestanian, Phys. Rev. Lett. \textbf{108}, 038303 (2012)

\bibitem{supplmat} Details regarding the method of solution are given in the
Supplementary Materials.

\bibitem{Fay08} S. Fayolle, T. Bickel, A. W\"{u}rger, Phys. Rev. E \textbf{77%
}, 042404 (2008)

\bibitem{Wue10} A. W\"{u}rger, Rep. Prog. Phys., \textbf{73}, 126601 (2010)

\bibitem{Der87} B. Derjaguin, N. Churaev, V. Muller, \textit{Surface Forces}
(Plenum, New York, 1987)

\bibitem{Bla71} J.R. Blake, J. Fluid Mech. \textbf{46}, 199 (1971)

\bibitem{Ruc81} E. Ruckenstein, J. Coll. Interf. Sci., \textbf{83}, 77 (1981)

\bibitem{Put05} S.A. Putnam, D.G. Cahill., Langmuir \textbf{21}, 5317 (2005)

\bibitem{Wue08} A. W\"{u}rger, Phys. Rev. Lett., \textbf{101}, 108302 (2008)

\bibitem{Vig10} D. Vigolo, S. Buzzaccaro and R. Piazza, Langmuir, \textbf{26}%
, 7792 (2010)

\bibitem{Jia09} H.-R. Jiang, H. Wada, N. Yoshinaga and M. Sano, Phys. Rev.
Lett., \textbf{102}, 208301 (2009)

\bibitem{Bre61} H. Brenner, Chem. Eng. Sci. \textbf{16}, 242 (1961)

\bibitem{Morth10} J. Morthomas, A. W\"{u}rger, Phys.\ Rev. E \textbf{81},
051405 (2010)

\bibitem{The12} I. Theurkauff, C. Cottin-Bizonne, J. Palacci, C. Ybert, L.
Bocquet, Phys. Rev. Lett.\textbf{\ 108}, 268303 (2012)






\end{thebibliography}
\end{document}